\documentclass[showpacs,floats,floatfix,aps,prl,groupedaddress,amsfonts,amssymb,twocolumn]{revtex4}
\usepackage{graphicx}
\usepackage{psfrag}
\usepackage{psfig}
\usepackage{amsmath}
\newcommand{\be}[0]{\begin{eqnarray}}
\newcommand{\ee}[0]{\end{eqnarray}}

\begin{document}

\title{Make life simple: unleash the full power of the parallel tempering algorithm}
\author{Elmar Bittner}
\author{Andreas Nu{\ss}baumer}
\author{Wolfhard Janke}
\affiliation{Institut f\"ur Theoretische Physik and Centre for
Theoretical Sciences (NTZ),\\
Universit\"at Leipzig, Postfach 100\,920, D-04009 Leipzig, Germany}

\date{\today}

\begin{abstract}
\noindent
We introduce a new update scheme to systematically improve the efficiency
of parallel tempering
simulations. We show that by adapting the
number of sweeps between replica exchanges to the canonical autocorrelation
time, the average round-trip time of a replica in temperature space can be
significantly decreased. The temperatures are not dynamically adjusted as
in previous attempts but chosen to yield a $50\%$ exchange rate of adjacent
replicas. We illustrate the new algorithm with results for the Ising model
in two and the Edwards-Anderson Ising spin glass in three dimensions.
\end{abstract}

\maketitle

The parallel tempering (PT), or replica exchange, simulation technique~\cite{pt0,pt1,pt2,pt3}
provides an efficient method to investigate systems with rugged free-energy landscapes~\cite{janke},
particularly at low temperatures. 
Initially, applications of the method were limited to problems in statistical 
physics. By now, however, PT and its extensions are used in many disciplines,
e.g.\ biomolecules~\cite{proteins1,proteins2,proteins3,proteins4,proteins5,proteins6,proteins7}, 
bio\-informatics~\cite{bioinfo}, 
zeolite structure solution~\cite{zeolit},
classical and quantum frustrated spin systems~\cite{frustrated1,frustrated2},
spin glasses~\cite{pt2,pt3,spinglass2,spinglass3,spinglass4,spinglass5} and
QCD~\cite{hep1,hep2,hep3}.
The use of PT in interdisciplinary fields spanning physics, chemistry, biology, engineering
and material sciences rapidly increases. 

In a PT simulation, one generates many replicas of Monte Carlo (MC) Markov chains 
or molecular dynamics (MD) trajectories
at different 
temperatures in parallel. At regular intervals an attempt is made to exchange the 
configurations of different, usually adjacent replicas,
which is accepted with probability 
\begin{equation}
		  \label{prob}
		  P_{\rm PT}(E_1,\beta_1 \rightarrow E_2,\beta_2)=\min [1,\exp(\Delta \beta \Delta E)],
 \end{equation}
where $\Delta \beta=\beta_2-\beta_1$ is the difference between the inverse temperatures of the two replicas and
$\Delta E=E_2-E_1$ their energy difference. The acceptance probability is the smaller
the larger the temperature difference or the system size gets. 
For PT simulations to be most efficient, each replica should spend the same amount of time
at each temperature. To this end, several strategies have been proposed in the last 
years~\cite{opti1,opti2,opti3,opti4,opti5,opti6,opti7}, but an efficient selection of optimal 
temperature intervals is still an open problem.
In the physically appealing protocol proposed by Katzgraber {\it et al.}~\cite{opti5}
the optimal temperatures are determined from the 
flow in temperature space, that is, 
the rate of round trips between low and high temperatures is maximized 
by systematically re-adjusting the temperatures. 

Unfortunately, their initial recursion is rather complex and needs a 
significant amount of CPU time. 
Therefore, we do not use the idea of maximum flow and rather employ the concept  
of a constant acceptance rate between adjacent replicas, which can be calculated
from
\begin{equation}
		  \label{acc}
		  A(1 \rightarrow 2)=\!\!\sum_{E_1,E_2} P_{\beta_1}(E_1) P_{\beta_2}(E_2) 
		   P_{\rm PT}(E_1,\beta_1 \rightarrow E_2,\beta_2),
 \end{equation}
where $P_{\beta_i}(E_i)$ is the probability for replica $i$ at $\beta_i$ to have the energy $E_i$
(the subscript is the replica index).
Using this formula we can calculate, starting from $\beta_1$, a set of inverse temperatures $\beta_i$
which satisfy $A(i \rightarrow i+1)={\rm const.}$.
For systems with a diverging specific heat one obtains 
a high density of replicas around the critical temperature, i.e., 
the difference between the inverse temperature of two adjacent
replicas is small.
For high values of $\beta$, i.e.\ low temperatures, the difference between energy distributions at different 
temperatures becomes small and therefore $\Delta \beta$ increases. Furthermore for small $\beta$ values, $\Delta E$ 
decreases and the spacing between the replicas grows. 

\begin{figure}[t]
		  \centerline{
		  \includegraphics[scale=0.6]{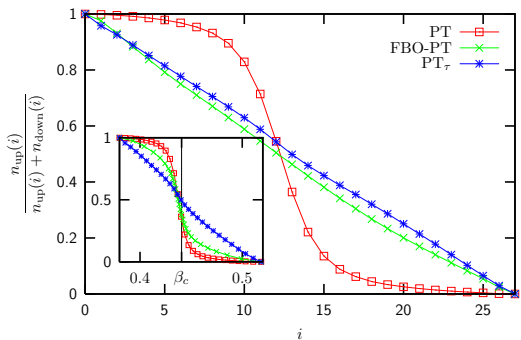}
		  }
		  \caption{
		  \label{fig_flow}
		  Fraction of replicas which wander from the smallest $\beta_i$ to the largest
		  as a function of the replica index $i$ for the 2D Ising model ($L=80$).
  The simulations without optimization exhibit a sharp decline 
  close to $\beta_\text c$, as one can see in the inset. 
  Taking the 
  canonical correlation times $\tau_{\rm can}$ into account (PT$_\tau$), the fraction 
  decreases, for the same set of temperatures, almost linearly.
		  }
 \end{figure}

As an illustration, we shall first consider 
MC simulations of 
the two-dimensional (2D) Ising model
where the density of states and hence (\ref{acc}) can be calculated exactly~\cite{beale}.
For all reasonably chosen rates $A(1 \rightarrow 2)$, the replica flow from 
high to low temperatures and vice versa turns out to be very slow, at least
when a local update scheme, e.g.\ the Metropolis algorithm, is used for 
each of the replicas. The replica flow through the temperature space
shows a significant drop around the critical temperature. In 
Fig.~\ref{fig_flow} we show as an example for an acceptance rate of $50\%$
the fraction of replicas which have visited most recently the smallest $\beta$
value and wander ``up'' in the inverse temperature space. This sharp drop-off 
behavior led Katzgraber {\it et al.}~\cite{opti5} to their feedback-optimized
update scheme (FBO-PT), in which they re-adjust the temperatures by analyzing
the local diffusivity.

\begin{figure}[t]
        \centerline{
		  \includegraphics[scale=0.6]{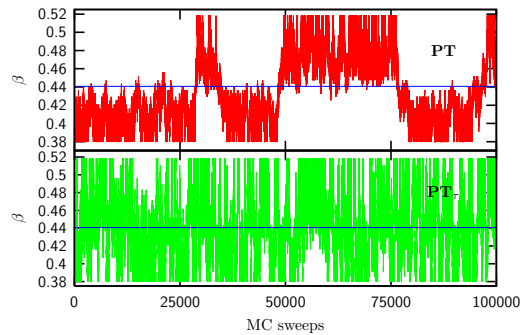}
        }
        \caption{
        \label{fig_block}
		Time series of an arbitrarily chosen replica on its way through inverse
	    temperature space of the 2D Ising model ($L=80$). The upper plot shows
	    the result of a PT simulation and the lower that of a
	    PT$_\tau$ simulation with $N_{\rm local}=\tau_{\rm can}$.  The
	    horizontal lines indicate $\beta_c$, the infinite-volume critical
	    point. The blocks in the time series are
	    a signal of the increasing autocorrelation times due to critical
	    slowing down.  
	} 
\end{figure}

We, on the other hand, want to remove the unwanted behavior at $\beta_c$, while
keeping the temperatures fixed at their initial values.
Looking at the trajectory of an arbitrarily chosen replica in temperature space shown in the
upper plot of Fig.~\ref{fig_block}, we see a
clear block structure, where the border of the blocks coincides with the critical temperature.
Such a block structure is related to a bottleneck in the flow through the temperature space, or, 
in other words, for a replica starting from a high temperature it is hard to overcome 
this bottleneck and move to the low temperature region.
A plausible explanation of this observation is as follows: toward the critical temperature
the autocorrelation time increases due to critical slowing down 
and therefore two exchanged replicas stay in phase space close to each other. It is 
hence more likely that these two replicas exchange again.

To verify this idea, we use a toy model based on the 
bivariate Gaussian process with $0 \le \rho < 1$~\cite{gauss_process},
\begin{equation}
\label{gauss}
		  e_i=\rho e_{i-1}+\sqrt{1-\rho^2}e_i'~,\quad i\ge 1~,
\end{equation}
where $e_0=e_0'$, and the $e_i'$ 
are {\it independent} Gaussian random variables satisfying 
$\langle e_i'\rangle=0$ and $\langle e_i' e_j'\rangle-\langle e_i'\rangle^2=\delta_{ij}$. 
Iterating this recursion it follows that the 
autocorrelation function is
$		  A(k)=\langle e_0 e_k\rangle=\rho^k\equiv e^{-k/\tau_{\rm exp}}~,$
where $\tau_{\rm exp}=-1/\ln \rho$ is the exponential autocorrelation time. 
It can be shown that with increasing
$\tau_{\rm exp}$ the mean step size decreases, i.e., 
$\langle | e_{i+1}-e_i|\rangle = 2 \sqrt{(1-\rho)/\pi}$, such 
that the system moves slower through the one-dimensional phase space, and 
this is what we are interested in.

Using the stochastic process (\ref{gauss}) we are able to approximate for any realistic 
model the movement in energy space during a parallel tempering simulation. 
From the energy distribution of initial canonical simulations
we obtain for each of the replica at $\beta_i$ 
the mean and variance which, after a trivial shift and
rescaling, can be reproduced with (\ref{gauss}).   
Next we exploit the freedom in the model
to adjust $\tau_{\rm exp}$ for each temperature which allows us to
investigate the dependence of the flow through temperature space on the autocorrelation times.
In general, simulations near a second-order phase transition are affected by critical slowing down, i.e., an
increasing autocorrelation time $\tau_{\rm can} \sim\xi^z$, where $\xi$ denotes the (spatial) correlation length 
and $z$ is the dynamical critical exponent. 
To take this into account, we set $\tau_{\rm exp}$ to the canonical 
autocorrelation time $\tau_{\rm can}$ of the energy measured in the independent 
simulations.
Together with the mean and variance this specifies the parameters
of the replicated process (\ref{gauss}).

By fitting to 2D Ising model MC data,
our first finding comes from a comparison of the autocorrelation times
for iterations of (\ref{gauss}) with and without the PT routine.  As expected, the autocorrelation times for the PT simulation are much smaller.
The flow through temperature space looks similar as for the 2D Ising model 
depicted in Fig.~\ref{fig_flow}. 
We also find a pronounced decline around the pseudo-critical point $\beta_c$.
The reason for this behavior is, as already anticipated above, the 
slowed down dynamics near $\beta_c$. That means, after two adjacent replicas in the vicinity of 
$\beta_c$ have been exchanged, they will stay close to each other and changing them back to the original state 
is more likely than an exchange with another replica. 
If the dynamics is even slower (by simply tuning $\tau_{\exp}$ larger)
a complete trapping can be observed and the replicas
do not move from low to high temperatures at all. 

By systematically varying the inputted autocorrelation times, 
our toy model suggests that an easy way to cure this problem is to increase the number of local 
updates between the PT exchanges proportional to the autocorrelation time of the initial (non PT) 
simulation for a given temperature.
\begin{figure}
        \centerline{
		  \includegraphics[scale=0.6]{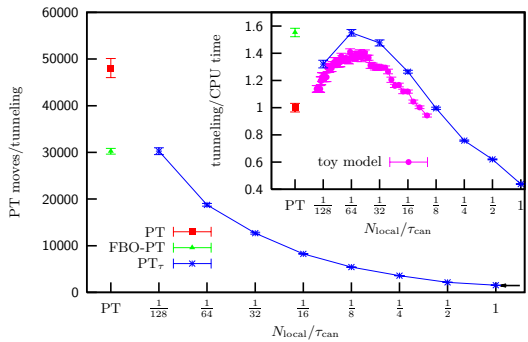}
        }
        \caption{
        \label{fig_tunneling}
		  PT moves per tunneling as a function of $N_{\rm local}$ for 
		  the 2D Ising model ($L=80$), approaching for 
		  $N_{\rm local} \approx \tau_{\rm can}$
		  the unbiased random walk limit. 
		  The inset shows the actually needed computing time in 
		  units of the total run time of the standard PT simulation. 
        }
 \end{figure}

This general strategy will be now first tested for the 2D Ising model. 
For system sizes up to $L=80$ we use the exact energy distributions~\cite{beale} to calculate
a set of inverse temperatures $\{\beta_i\}$ with an acceptance rate $A(i\rightarrow i+1)=0.5$ starting from $\beta_1=0.38$. 
To cover almost the same temperature interval for different system sizes $L$ the number of replicas $N$ has to
increase with $L$~\cite{ebwj_clus}. For
this set of inverse temperatures we perform short independent Metropolis MC simulations
to estimate the canonical autocorrelation times $\tau_{\rm can}(\beta)$ of the energy,
together with the mean and width of the energy distribution.
In the actual simulations we then use the usual PT update scheme with 
only one important modification, namely, we
choose the number of sweeps $N_{\rm local}(\beta)$
between the attempts to exchange the configurations proportional 
to $\tau_{\rm can}(\beta)$
($N_{\rm local}(\beta)=1$ for standard PT and FBO-PT simulations).
The larger the number of sweeps between the exchange attempts, the smaller 
the correlation between adjacent replicas. Therefore, one has to find a 
compromise between accuracy and computer time, which can be easily
achieved by using our toy model (which runs orders of magnitude faster than the
actual simulations).
To illustrate this we include in Fig.~\ref{fig_tunneling} a comparison for different 
choices of $N_{\rm local}$. 

The main plot of Fig.~\ref{fig_tunneling} shows the number of PT moves  
necessary for an arbitrarily chosen replica to move from the highest to the lowest temperature and back again.
In the following such a round trip will be called tunneling. 
We clearly see that with increasing
number of sweeps per replica the tunneling time converges to the value of an unbiased random walk (indicated 
by the arrow in the lower right corner)
consisting of two legs of length $(N-1)$. The limit for one round trip is hence
given by $2 (N-1)^2$. If we choose $N_{\rm local}(\beta)=\tau_{\rm can}(\beta)$,
the correlation
between adjacent replicas is negligible and each replica performs a random walk through temperature
space (see lower plot in Fig.~\ref{fig_block}). Furthermore, the sweeps needed for a tunneling event
are close to the theoretical value,
as is also reflected in the inset of Fig.~\ref{fig_flow}, where we show that the fraction of replicas moving ``up'' in the inverse 
temperature is an almost linear function of $\beta$.
This is a major difference to FBO-PT~\cite{opti5}, where the linear relation holds 
only for the fraction of replicas moving ``up'' as a function of the {\it replica index}.
In the inset of Fig.~\ref{fig_tunneling} we compare the computational cost 
of our improved PT (denoted by PT$_\tau$)
with that for standard PT and FBO-PT, showing
that for moderate values of $N_{\rm local}$ the computational effort is the same for both improved methods.
To keep the comparison fair, we have excluded the additional computer time needed for FBO-PT 
to determine the set of inverse temperatures and for PT$_\tau$ to obtain the local autocorrelation times.
If one increases $N_{\rm local}$, the ratio of tunnelings per CPU time decreases, i.e., above a certain threshold value of $N_{\rm local}$
the computational effort of PT$_\tau$ increases faster than the improvement of the tunneling speed. 

\begin{figure}[t]
        \centerline{
                  \includegraphics[scale=0.6]{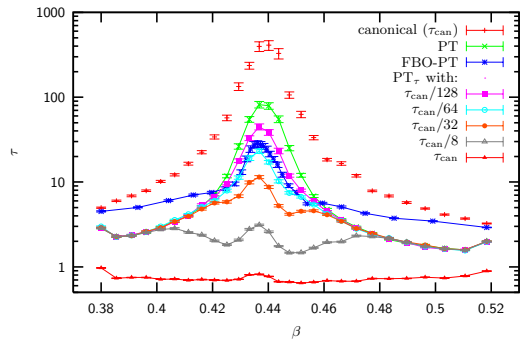}
        }
        \caption{ 
        \label{fig_tau}
                   Autocorrelation times as function of inverse temperature for the canonical
                   simulations, the standard PT update scheme, the feedback-optimized PT
                   method (FBO-PT), and four different runs of our improved PT update 
                   scheme (PT$_\tau$) for the 2D Ising model ($L=80$).
        }
 \end{figure} 

To compare our improved PT$_\tau$ with other 
methods one should not only look at the computational cost but also at the accuracy that is achieved 
for the same amount of measurements.
An easy way to check this is to measure the autocorrelation time $\tau$.
In Fig.~\ref{fig_tau} we show the autocorrelation times of the 2D Ising model with $L=80$ for standard PT,
FBO-PT, and our PT$_\tau$ algorithm with different choices of $N_{\rm local}(\beta)$.  
The improvement gained by using PT instead of simulating each temperature independently is almost one order of magnitude in the
region around the critical point. If one rearranges the inverse temperatures according to the FBO-PT algorithm one
finds even smaller autocorrelation times around $\beta_c$, but the improvement away from criticality is less pronounced than for
the standard PT method. Taking in PT$_\tau$ the local autocorrelation times 
$\tau_{\rm can}(\beta)$ into account we can 
decrease $\tau$ systematically.
For $N_{\rm local}(\beta)=\tau_{\rm can}(\beta)/64$, where for all temperatures the autocorrelation times of the PT$_\tau$ simulation are slightly smaller
than for FBO-PT, the computational effort is almost equal for the two methods. 
If we use $N_{\rm local}(\beta)=\tau_{\rm can}(\beta)$, then the autocorrelation times are smaller than unity for all temperatures
and the resulting time series are nearly uncorrelated, but the computational costs are clearly too high to make this choice useful.

\begin{figure}
        \centerline{
		  \includegraphics[scale=0.6]{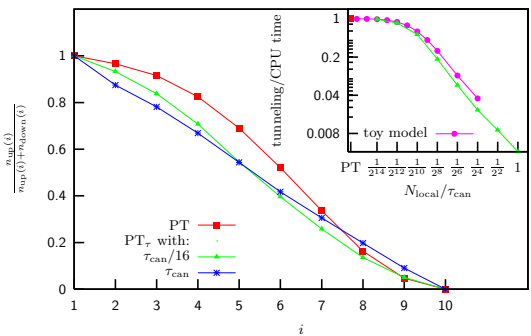}
        }
        \caption{
        \label{fig:eai_eta_tunnel}
		 Fraction of replicas which wander from the smallest $\beta_i$ to the
	    largest as a function of the replica index $i$ for the 3D
	    Edwards-Anderson Ising spin-glass model ($L=6$, averaged over $20$ disorder realizations).
        }
 \end{figure}

We close with a brief remark on applications of our PT$_\tau$ algorithm
to a MC study of
the 3D Edwards-Anderson Ising spin-glass model on a $L=6^3$ lattice simulated in
a temperature range from $0.75$ to $1.7$ around $T_c\sim1.15$. Using the same procedure as
described above,
we find also here an improvement of the replica flow from high to low temperatures, i.e., from 
the disordered to the spin-glass phase (see Fig.~\ref{fig:eai_eta_tunnel}). However, the 
additional computational effort to gain
this improvement is significant due to the exponential increase of the autocorrelation time
with decreasing temperature. Therefore, one has to carefully tune the balance between used computer time and
quality of results.

To summarize, we discovered a remarkable block building structure in 
PT simulations, revealed the mechanism behind it and showed how to cure 
this problem by taking into account the temperature dependence of 
autocorrelation times. This demonstrates how easily the quality 
of PT simulation data can be improved both in MC and MD studies. 

Work supported by the Deutsche Forschungsgemeinschaft (DFG)
under grant Nos.~JA483/22-1 and JA483/23-1, 
the EU RTN-Network `ENRAGE' 
under grant
No.~MRTN-CT-2004-005616, and by the computer-time grant No.~hlz10 of 
NIC J\"ulich.

\end{document}